\documentclass[a4paper, 10pt]{article}
\usepackage{pos}
\usepackage{physics}
\usepackage{graphicx}
\usepackage{float}

\newcommand{\V}{\mathcal{V}}
\newcommand{\M}{\mathcal{M}}
\renewcommand{\S}{\mathcal{S}}

\newcommand{\K}{\mathcal{K}}

\title{Growth of Cosmic Strings with Field-Dependent Tension}

\author*[a, b]{Luca Brunelli}

\affiliation[a]{Department of Physics and Astronomy, Bologna University \\
Via Irnerio 46, 40126, Bologna, Italy}

\affiliation[b]{INFN, Sezione di Bologna, viale Berti Pichat 6/2, 40127 Bologna, Italy}

\emailAdd{l.brunelli@unibo.it}

\abstract{A novel mechanism to produce a cosmic network of fundamental superstrings based on a time-varying string tension has been recently proposed. It has been found that fundamental superstrings can grow in a kinating background driven by the rolling of the volume modulus of Type IIB string compactifications towards the minimum of its potential. In this talk, I will generalise this analysis using dynamical systems techniques. First, I will analyse the cosmological growth of strings with a field-dependent tension in a Universe filled with a perfect fluid, finding a \emph{growth region} in the phase space of this system. This machinery is then applied to both fundamental superstrings and effective strings obtained from wrapping $p$-branes on $(p-1)$- dimensional cycles. I will show how cosmological growth can be achieved in both cases not only in kinating backgrounds, but also in scaling fixed points. This talk is based on \cite{Brunelli:2025ems}.}

\FullConference{3rd General Meeting of the COST Action COSMIC WSIPers (CA21106) (COSMICWISPers2025)
                3rd Training School of the COST Action COSMIC WSIPers (CA21106) (COSMICWISPers2025)\\
9–12 Sept 2025 and 16–19 Sept 2025\\
Sofia, Bulgaria and Annecy, France\\}

\begin{document}

\maketitle

\section{Introduction}

Cosmic strings are one-dimensional objects emerging in a variety of BSM theories \cite{Vilenkin:2000jqa}. A network of cosmic strings is particularly interesting, as its energy density tends to reach a scaling regime with respect to a background fluid \cite{Martins:2000cs, Revello:2024gwa}. In fact, since networks are a continuous source of gravitational waves (GWs), they can offer insights into the cosmological history of our Universe.

In string theory, networks of cosmic strings can form as a consequence of very energetic phenomena, like the annihilation process at the end of brane-antibrane inflation (for a recent perspective see \cite{Cicoli:2024bwq}). Alternatively, the authors of \cite{Conlon:2024uob} proposed a new mechanism for the formation of a cosmic network of superstrings: an initial population of loops of fundamental strings can grow in comoving size as a consequence of their decreasing tension and eventually percolate.
The tension of fundamental (F-) strings in Type IIB depends on the vacuum expectation value (vev) of the volume modulus, and if this field experienced a period of kination in the early Universe, F-strings can grow to cosmic sizes. 
In this talk we generalise this framework to include an arbitrary perfect fluid and show that the comoving growth of cosmic superstrings with field-dependent tension can occur not only in kination, but also at stable scaling fixed points

\section{Strings with Field-Dependent Tension}

Varying the Nambu-Goto action with a time-dependent tension $\mu(t)$ in a FLRW spacetime with scale factor $a(t)$, we can retrieve two equations of motion for circular loops of strings with a time-dependent radius $R(t)$. One of the two is particularly enlightening:
\begin{equation}\label{eq:eom string}
    \frac{\dot \epsilon}{\epsilon} = H-a^2 \dot R^2 \left(2 H + \frac{\dot \mu}{\mu}\right)
\end{equation}
where $\epsilon = a R_{\rm max}$ is the proper length of the string. Clearly, comoving growth happens when $\dot \epsilon/\epsilon >H$ which, from \eqref{eq:eom string} is realised when the following Growth Condition (GC) holds:
\begin{equation}\label{eq:gc}
    2H+ \frac{\dot\mu}{\mu} <0.
\end{equation}
We will be interested in the case in which the time-dependence of the string tension is set by the evolution of a scalar field $\varphi(t)$. In particular, we will consider an exponential dependence on $\varphi(t)$:
\begin{equation}\label{eq:exp mu}
    \mu(t) = \mu_0 \, e ^{- \xi \frac{\varphi(t)}{M_p}}.
\end{equation}
By means of \eqref{eq:gc} and \eqref{eq:exp mu} we can directly relate the dynamics of $\varphi$ with the growth of the string. In order to do so, we shall employ dynamical system techniques. First of all, we define the phase space variables:
\begin{equation}
    X^2 = \frac{\dot \varphi^2}{6 H^2 M_p^2}, \quad Y^2 =   \frac{V(\varphi)}{3 H^2 M_p^2}
\end{equation}
where $V(\varphi)$ is the potential $\varphi$ is subject to, we will assume an exponential form:
\begin{equation} \label{eq:exp V}
    V(\varphi) = V_0 \, e^{-\lambda \frac{\varphi}{M_p}}
\end{equation}
Together with the field, we shall consider a background perfect fluid with equation of state $p = \omega \rho$.
The Friedman equation:
\begin{equation}\label{eq:friedman}
    X^2 + Y^2 + \frac{\rho}{3 H^2 M_p^2} = 1
\end{equation}
implies $X^2+Y^2 \leq 1$, and since we restrict ourselves to expanding Universes ($H>0$), the phase space is the unit semi-circle in the upper XY plane. 
We can then rewrite the Klein-Gordon equation for the scalar field and the continuity equation for the background fluid in terms of $X$ and $Y$ as an autonomous system:
\begin{align}
    X' & = -3X + \lambda \sqrt{\frac{3}{2}} Y^2+ \frac{3}{2}X\left[ 2X^2 + (\omega +1) (1-X^2-Y^2)\right] \label{eq:X'}\\
Y' & = - \lambda \sqrt{\frac{3}{2}}\, XY + \frac{3}{2} Y \left[ 2X^2 + (\omega +1) (1-X^2-Y^2)\right], \label{eq:Y'}
\end{align}
where derivatives are taken with respect to the number of efolds $N = \ln a$. The fixed points of the dynamical system \eqref{eq:X'}-\eqref{eq:Y'} have been fully classified in \cite{Copeland:1997et}. 
The GC \eqref{eq:gc} identifies a Growth Region (GR) in the phase space:
\begin{equation}
    \begin{cases}
        X> \frac{1}{\xi}\sqrt{\frac{2}{3}} \; \text{ if } \xi >0\\
         X< \frac{1}{\xi}\sqrt{\frac{2}{3}} \; \text{ if } \xi <0\, .
    \end{cases}
\end{equation}
If the system spends time in the GR, the string grows. Since $|X| \leq 1$, the GR only exists if:
\begin{equation}\label{eq:lower bound xi}
    |\xi| > \sqrt{\frac{2}{3}}. 
\end{equation}
The field dominated  and the scaling fixed points of the dynamical system \eqref{eq:X'}-\eqref{eq:Y'} ($\M$ and $\S$ respectively) can be in the GR (provided it exists) depending on  $\lambda$ and $\omega$.  On the other hand, one of the kinating fixed points $\K_\pm = (\pm 1, 0)$ will be within the GR.  Let us now apply this machinery to F-strings of Type IIB string theory.

\section{Fundamental Strings in Type IIB String Theory}

F-strings are the building blocks of string theory. Their tension is naturally linked to the string scale $M_s^2$, which in 4d Einstein frame can be expressed as:
\begin{equation}\label{eq:f-tension}
    \mu \simeq M_s^2 = \frac{\sqrt{g_s}}{4 \pi \V} M_p^2
\end{equation}
where $g_s$ is the string coupling, controlled by the dilaton field, and $\V$ is the volume of the compactification manifold in string units. Let us consider the dependence of $\mu$ on the volume modulus. Upon canonical normalisation : $\V = \exp\left(\sqrt{3/2}\, \Phi/M_p\right)$, the tension becomes exponential in $\Phi$:
\begin{equation} \label{eq:mu Phi}
    \mu \sim M_p^2 \, e^{-\sqrt{\frac{3}{2}} \frac{\Phi}{M_p}}\,.
\end{equation}
Comparing this with \eqref{eq:exp mu} yields:
\begin{equation}\label{eq:xi volume}
    \xi =\sqrt{\frac{3}{2}} >\sqrt{\frac{2}{3}}
\end{equation}
so a GR exists for the volume, corresponding to the $X>2/3$ section of its phase space. 
Now, the typical form of the scalar potential of the volume far away from its minimum is an inverse power, that upon canonical normalisation becomes of the form \eqref{eq:exp V}. This allows us to use our DS techniques and find under what conditions the fixed points of the system lie in the GR. 

In this case, $\K_+ $ is clearly in the GR. This is precisely the result of \cite{Conlon:2024uob}: a volume kinating towards large values leads to comoving growth. However, $\K_+$ is unstable, and as soon as the system is displaced, it will move away from it. Depending on how close to $\K_+$ the system initially is, it can spend quite a long time in the GR before reaching its final attractor.  
The field-dominated fixed point $\M = \left(\lambda/\sqrt{6}, \sqrt{1- \lambda^2/6}\right)$ might also lie in the GR.  Combining its existence condition found in \cite{Copeland:1997et} with $X>2/3$ we get:
\begin{equation}
    2 \sqrt{\frac{2}{3}}< \lambda < \sqrt{6}. 
\end{equation}
For these values of $\lambda$, also $\M$ is unstable. The scaling fixed point $\S = \left(\sqrt{\frac{3}{2}}\frac{\omega + 1}{\lambda}, \sqrt{\frac{3(1-\omega^2)}{2 \lambda^2}}\right)$, instead, is always stable when it exists. Therefore, the system can spend a long time in the GR if $\S$ is inside. The combined existence and growth conditions  for $\S$ yield:
\begin{equation}
    \sqrt{3(\omega+1)} < \lambda < (3/2)^{3/2}(\omega+1)
\end{equation}
and this window in parameter space exists as long as $\omega >-1/9$.  

\section{Effective Strings in Type IIB String Theory}

A different type of stringy object in the 4d EFT are \emph{effective strings}: $p-$branes wrapping $(p-1)$-cycles in the internal dimensions. If the wrapped cycle has volume $t_{p-1}$ in string units, the tension acquires an additional moduli-dependence:
\begin{equation}\label{eq:mu effective}
    \mu \sim M_s^2\,  t_{p-1}. 
\end{equation}
In Type IIB, the only BPS-stable objects one can build this way that display interesting dynamics are D3-strings (i.e. D3-branes wrapped around 2-cycles) and NS5-strings (NS5-branes wrapped around 4-cycles).
As pointed out in \cite{Conlon:2024uob}, for the simplest realisation of the volume, $\V = t^3 = \tau^{3/2}$, these do not feature any GR, since $\xi_{D3} = \sqrt{2/3}$ and $\xi_{NS5}= \sqrt{1/3}$, both below the bound \eqref{eq:lower bound xi}. Although the simplest volume does not allow for growth of effective strings, a fibred geometry introduces anisotropic moduli directions that modify the canonical normalisation. In fact, in a manifold defined as a $T^4$- or $K3$- fibration over a  $\mathbb P^1$ base, the volume acquires a new cycle dependence:
\begin{equation}
    \V =  k_{122}t_1 t_2^2 = \kappa \tau_b \sqrt{\tau_f} 
\end{equation}
where $k_{122}$ is an integer triple intersection number, $t_1$ and $\tau_b$ are the base 2/4 cycle volumes respectively, and $t_2$ and $\tau_f$ are the fibre 2/4 cycle volumes respectively. The idea is to move in field space in the direction orthogonal to $\V$, namely: $u \propto \tau_f/\tau_b \sim \exp(\sqrt{3}\, \phi/M_p)$, where $\phi$ is the canonical normalisation of $u$. 
Since the tension \eqref{eq:mu effective} is linear in the volume of the cycle wrapped by the brane, we just have to check the canonical normalisation of each cycle to see whether there exists a GR in their phase space. One finds that:
\begin{equation}
    \tau_f  \sim \V^{2/3} e^{\frac{2}{\sqrt 3} \frac{\phi}{M_p}},  \quad t_1 \sim \V^{1/3} e^{-\frac{2}{\sqrt 3} \frac{\phi}{M_p}}
\end{equation}
so that $|\xi| =2/\sqrt 3> \sqrt{2/3}$ for the base 2-cycle and fibre 4-cycle. On the other hand, for $t_2$ and $\tau_b$, $\xi$ does not satisfy the bound \eqref{eq:lower bound xi}. For the base D3-string, the GR is $X>1/\sqrt 2$, while for the fibre NS5-string $X<- 1/\sqrt 2$. Therefore, $\K_+$ and $\K_-$ are  within the respective GR's.   Moreover, the typical perturbative contribution to the potential is (inverse) monomial in $\tau_f$, so that, upon canonical normalisation, $V$ takes an exponential form akin to \eqref{eq:exp V}. Again, $\M$ and $\S$  can be in the GR of one field or the other. In general, we get that $\M$ is within one of the two GR's if:
\begin{equation}
    \sqrt{3}< |\lambda| < \sqrt{6}
\end{equation}
with the plus sign for $t_1$, minus sign for $\tau_f$. Once again, for these values of $\lambda$, $\M$ is unstable. The  scaling fixed point $\S$ lies in the GR if:
\begin{equation}
    \sqrt{3(\omega +1 )} < |\lambda| < \sqrt{3} (\omega+1)
\end{equation}
where this range of parameters exists for all $\omega >0$. As long as $\S$ exists, it is always stable. Therefore, once again we found stable orbits within the GR that can lead to long period of growth even for effective strings. 

\section{Conclusions and Outlook}

In this talk we showed that a population of string loops can experience comoving growth when their tension decreases, as happens when the tension depends on a compactification modulus. For fundamental strings, only the overall volume $\mathcal V$ can drive such growth. Even in scaling/tracking regimes, the tension can fall fast enough to sustain stable comoving growth while the field remains far from its minimum, where an exponential potential is a good approximation. This growth epoch can also occur for effective strings arising from 3/5-branes wrapping 2/4-cycles of a fibred CY.
In a follow-up work \cite{Brunelli:2025eif}, we studied the energy density of the string loop fluid as the field governing its tension approaches its minimum. We found that these strings can dominate the cosmic energy density, as in \cite{SanchezGonzalez:2025uco}, and that a fluid of NS5-strings can help resolve the \emph{overshoot problem} of steep stringy potentials.
When strings percolate into a network, one can compute the resulting GW spectrum, as done for F-strings in \cite{Ghoshal:2025tlk}. Extending this to effective strings is a natural next step. Another is to include fermionic worldsheet currents—ubiquitous in superstrings—which can modify the tension’s time dependence and potentially enhance growth. It would also be valuable to study the spectrum of light particles radiated by these cosmic (especially effective) superstrings and cross-correlate it with the GW spectrum, making them promising targets for advancing multi-messenger cosmology.

\section*{Acknowledgements}
I thank  M. Cicoli and F. G. Pedro for their collaboration in this project. I also thank F. G. Pedro and M. Cicoli for enlightening discussions about the present work. 
This article is based upon work from COST Action COSMIC WISPers CA21106,
supported by COST (European Cooperation in Science and Technology).

\bibliographystyle{utphys}
\bibliography{bibliography}

\end{document}